\documentclass{aa} 
\def\dosingle#1::::{#1}  \def\dodouble#1::::{ } 

\dodouble \documentclass[referee]{aa} ::::

\input epsfcopy.sty
\usepackage{natbib}

\renewcommand\citep[1]{(\citealt{#1})}

\newcommand\noarXiv[1]{#1} 

\newcommand\zzz[3]{#2}  

\def\la{\ifmmode\stackrel{<}{_{\sim}}\else$\stackrel{<}{_{\sim}}$\fi} 
\def\ga{\ifmmode\stackrel{>}{_{\sim}}\else$\stackrel{>}{_{\sim}}$\fi} 

\def\centreline{\centerline}
\def\hMpc{\mbox{$\,h^{-1}$ Mpc}}
\def\hMpcinv{\mbox{$\,h\,$Mpc$^{-1}$}}

\def\Omm{\Omega_{\mbox{\rm \small m}}}

\def\mydeg{\ifmmode^\circ\else$^\circ$\fi}

\def\llss{L_{\mbox{\rm \small LSS}}}

\def\rmax{r_{\mbox{\rm \small max}}}
\def\rsim{r_{\mbox{\rm \small sim}}}
\def\robs{r_{\mbox{\rm \small obs}}}
\def\anddd{{\mbox{\rm \ \ and\ \ }}}
\def\Psim{P^{\mbox{\rm \small sim}}}
\def\Prad{P^{\mbox{\rm \small rad}}}
\def\Ptan{P^{\mbox{\rm \small tan}}}

\def\SS{\S}  

\def\frtoday{Le\space\number\day\space\ifcase\month\or
  janvier\or f\'evrier\or mars\or avril\or mai\or juin\or
  juillet\or ao\^ut\or septembre\or octobre\or novembre\or 
  d\'ecembre\fi\space \number\year}

\def\tdefnsample{ 
\begin{table} 
\caption[]{The two angular subsamples of 
the quasar candidate catalogue \protect\citep{IovCS96}, defined by 
B1950 limits 
($\alpha_1\le \alpha\le \alpha_2$, 
$\delta_1\le \delta\le  \delta_2$)
and redshift limits $1.8 \le z < 2.4$. 
In \SS\protect\ref{s-corr3d},
these are further divided 
into redshift 
subsamples $1.8 \le z< 2.1$ and $2.1 \le z< 2.4$, 
yielding the four independent subsamples
whose analyses are shown in Figs~\protect\ref{f-corrna},
\protect\ref{f-corrnab} and \protect\ref{f-corrnb}.
In \SS\protect\ref{s-corrtanrad} (Fig.~\protect\ref{f-degen}), 
$1.8 \le z <2.4$ is used, i.e. yielding just two subsamples. 
When generating Poisson simulations for a given subsample, 
further subdivisions, at $\alpha'$ for 
the right ascension subsamples,
and at $\delta', \delta''$ for the declination subsamples, are 
defined in order to allow for variation in 
magnitude zero points or magnitude cutoffs between 
the different plates. That is, for a given subsample 
with several subdivisions 
$i$ defined by $\alpha'$ or $\delta',\delta''$,  
there are $n_i$ quasars in each 
subdivision $i$. For each subdivision $i$, 
$n_i$ uniform random $\alpha$ or $\delta$ 
values are generated in the appropriate $\alpha$ or $\delta$ range
for that subdivision.
This mimicks the magnitude limits. The subdivisions 
$\alpha'$, $\delta'$ and $\delta''$ are then ignored when calculating
$\xi$. Numbers of objects $N$ are indicated. 
\label{t-defnsample}} 
$$\begin{array}{cccc cccc} \zzz{\hline}{\hline}{} 
\alpha_1 & \alpha_2 & \delta_1 & \delta_2 &  \alpha' 
& \delta' & \delta'' & N\\ 
\multicolumn{8}{l}{\mbox{\rm `Right ascension ($\alpha$) subsample'}} \\ 
1^h00^m 
& 1^h59^m &   -42.0 & -37.5 & 1^h07^m 
& && 500\\ 
\multicolumn{8}{l}{\mbox{\rm `Declination ($\delta$) subsample'}} \\ 
0^h42^m & 1^h00^m &   -42.0 & -28.0 & 
& -37.5 & -32.5 & 453 \\ 
\zzz{\hline}{\hline}{} 
\end{array}$$ 
\end{table} 
}  

\def\tprobsim{ 
\begin{table} 
\caption[]{Probability (eq.~\protect\ref{e-defnpsim}) 
of finding one, two or three peaks among 200 random 
simulations at $f\llss$ where $f=0.5,1$ and/or $2$, 
for $(\Omm, \Omega_\Lambda)=$ $(0.4, 0.6)$. 
\label{t-probsim}} 
$$\begin{array}{ccc ccc c } \zzz{\hline}{\hline}{} 
\Psim_{0.5} & \Psim_{1} & \Psim_{2} & 
\Psim_{0.5,1} & \Psim_{1,2} & \Psim_{0.5,2} & \Psim_{0.5,1,2} \\ 
0.025 & 0.01 & 0.335 &  < 0.005 & 0.005 & 0.005 & < 0.005 \\ 
\zzz{\hline}{\hline}{} 
\end{array}$$ 
\end{table} 
}  

\def\fwedge{ 
\begin{figure} 
\centreline{\epsfxsize=70mm 
\zzz{\epsfbox[42 31 475 331]{"`gunzip -c H2494_1a.ps.gz"} } 
{\epsfbox[42 31 475 331]{"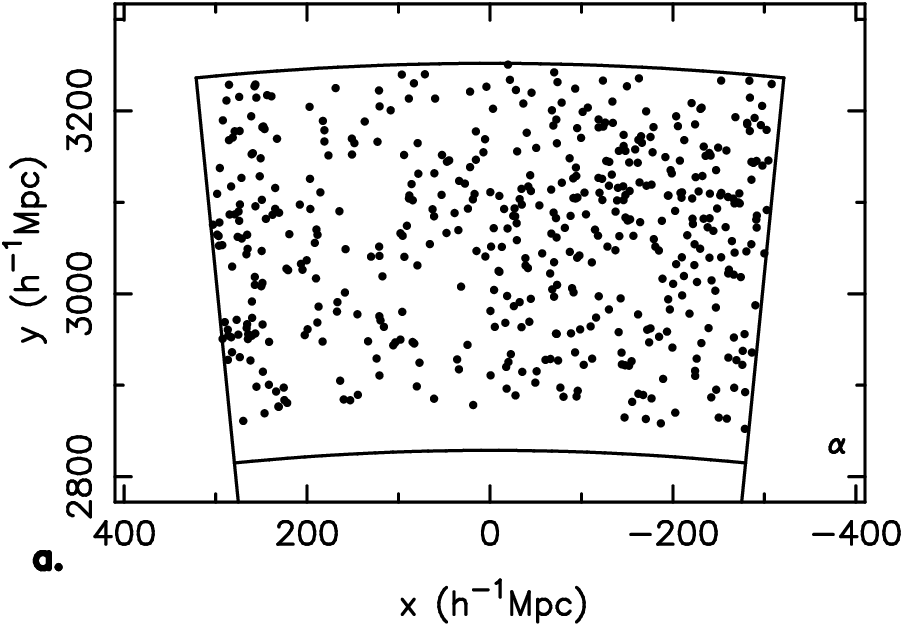"}}{}  } 
\centreline{\epsfxsize=70mm 
\zzz{\epsfbox[42 31 475 331]{"`gunzip -c H2494_1b.ps.gz"} } 
{\epsfbox[42 31 475 331]{"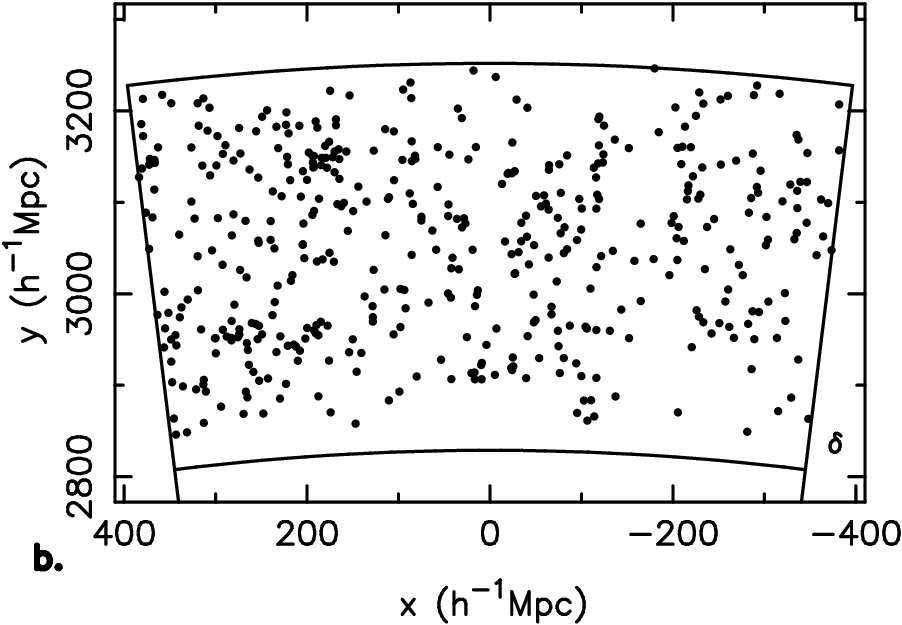"}}{}  } 
\caption[]{Projected comoving spatial distributions 
of the two quasar subsamples at redshifts $1.8 \le z < 2.4$, 
for $(\Omm=0.4, \Omega_\lambda=0.6)$. 
{\bf a.} The right ascension subsample, 
for $\alpha$ increasing from left to right. 
{\bf b.} The declination subsample 
for $\delta$ increasing from left to right. 
The discrete redshifts 
published \protect\citep{IovCS96} are converted to continuous 
values by uniform random offsets. The latter 
are used below. 
Differences in magnitude limits, hence different number densities, 
are visible in the $\alpha$ subsample and are corrected for 
in both subsamples (cf Table~\protect\ref{t-defnsample}). 
What appears to be a cluster or a supercluster in {\bf b.} 
at ($+200,3150$){\hMpc} is located at 
($\alpha\approx 1^h00^m, \delta\approx -38.5{\mydeg}, z\approx 2.24$). 
\label{f-wedge} 
} 
\end{figure} 
} 
 
\def\fcorrna{ 
\begin{figure}
\centreline{
 {\epsfxsize=70mm 
 \zzz{\epsfbox[56 40 531 530]{"`gunzip -c H2494_2.ps.gz"} } 
 {\epsfbox[56 40 531 530]{"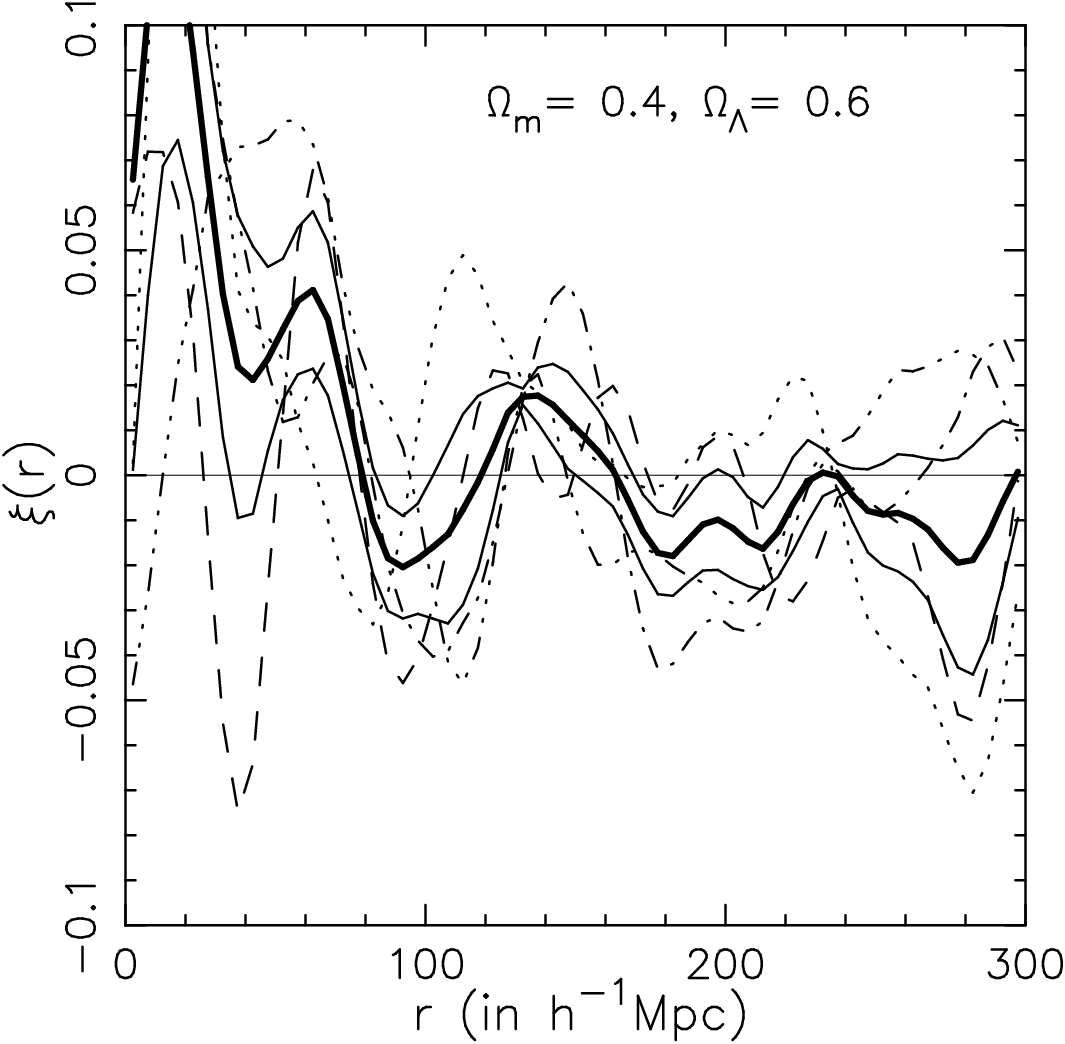"}}{}  }  
}
\caption[]{
Spatial two-point auto-correlation function 
$\xi(r)$, for separations $r$ in comoving units 
and ($\Omm=0.4,\Omega_\Lambda=0.6$). 
The four angular/redshift subsamples are shown as 
dashed ($\delta, 1.8\le z<2.1$), 
dashed-dotted ($\delta, 2.1\le z<2.4$), 
dotted ($\alpha, 1.8 \le z < 2.1$) and 
dashed-triple-dotted ($\alpha, 2.1 \le z < 2.4$) curves. 
The mean $\left<\xi\right>$ and the standard error in the mean 
$\sigma_{\left<\xi\right>}$ 
are shown by the thick and thin solid lines respectively. 
The correlation functions are calculated in three-dimensional 
curved space via 
$\xi(r) = (DD- 2DR/n +RR/n^2)/(RR/n^2)$ 
where $DD$, $DR$ and $RR$ indicate numbers of data-data, data-random 
and random-random quasar pairs respectively \protect\citep{LS93}, 
and $n=20$ times more 
random points than data points are used. 
The random catalogues use (i) uniform probability 
distributions in the two angular directions 
(Table~\protect\ref{t-defnsample}), and (ii) random permutations 
(``$z$ scrambles'', IIIb in ~\protect\citealt{Osmer81}) 
of the observational set of redshifts, to avoid 
biases from redshift selection effects \protect\citep{Scott91}. 
Bin size is $5${\hMpc} and 
$\xi$ is smoothed by a Gaussian with $\sigma=10${\hMpc}. 
The low values of $\xi$ at  $r\la 20${\hMpc} are related to redshift
roundoff error.
} 
\label{f-corrna} 
\end{figure} 
} 

\def\fcorrnab{ 
\begin{figure}
\centreline{
 {\epsfxsize=70mm 
 \zzz{\epsfbox[56 40 531 530]{"`gunzip -c H2494_3a.ps.gz"} } 
 {\epsfbox[56 40 531 530]{"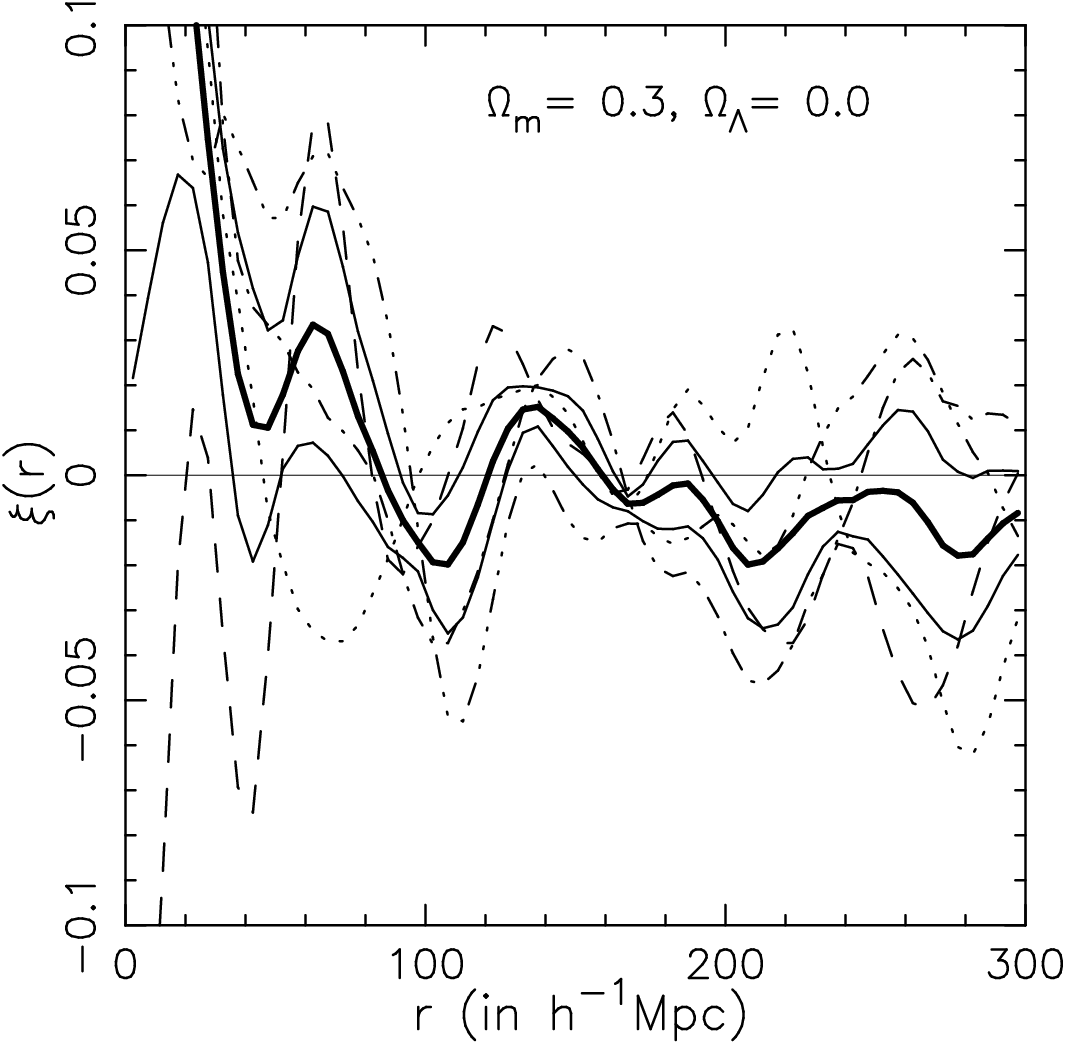"}}{}  }  }\par
\centreline{
 {\epsfxsize=70mm 
 \zzz{\epsfbox[56 40 531 530]{"`gunzip -c H2494_3b.ps.gz"} } 
 {\epsfbox[56 40 531 530]{"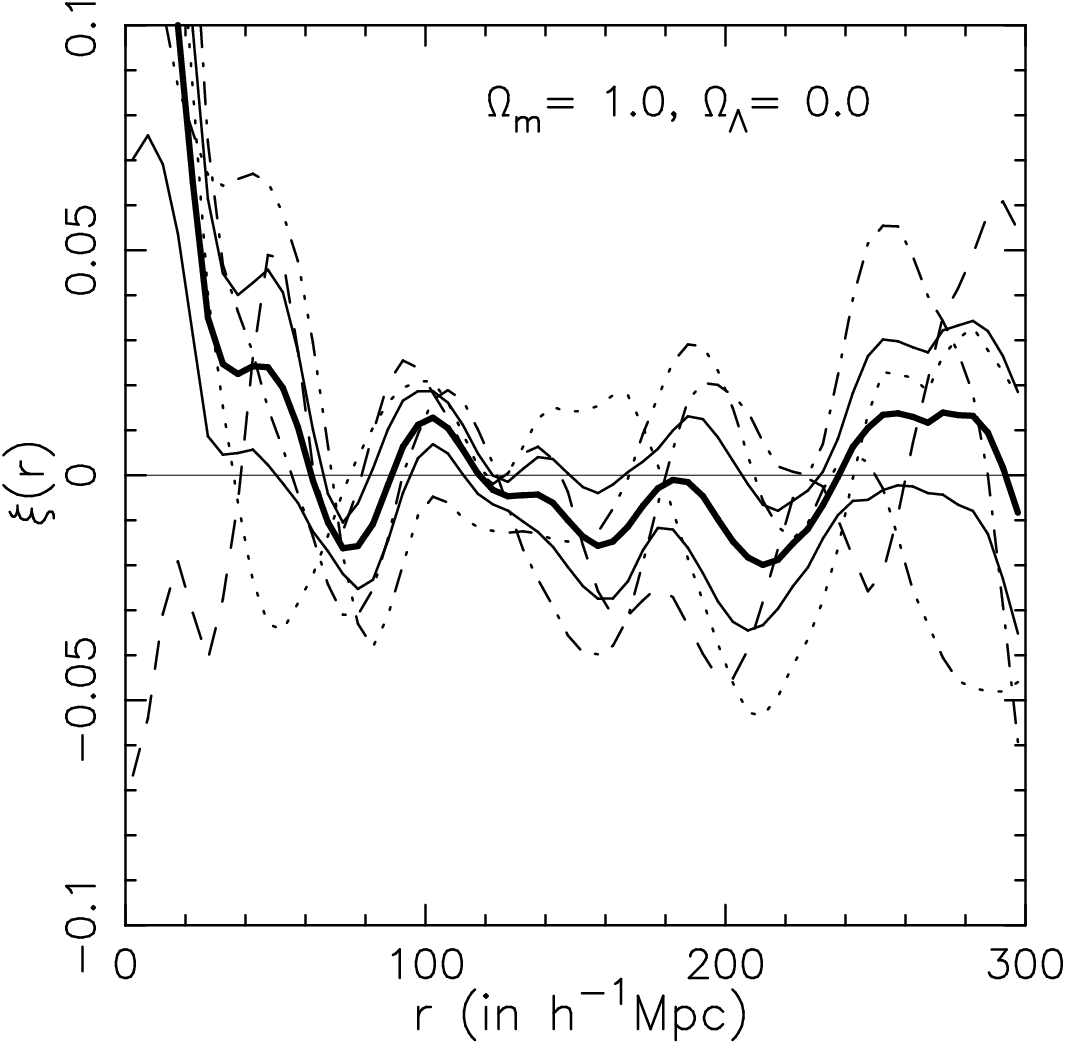"}}{}  }
}  
\caption[]{Correlation function $\xi(r)$, as for 
Fig.~\protect\ref{f-corrna}, but for a hyperbolic universe
$(\Omm=0.3,\Omega_\Lambda=0.0)$ above, and 
a zero cosmological constant, flat universe
$(\Omm=1.0,\Omega_\Lambda=0.0)$ below.
}
\label{f-corrnab}
\end{figure}
} 

\def\fcorrnb{ 
\begin{figure}
\centreline{
 {\epsfxsize=70mm 
\zzz{\epsfbox[56 78 459 482]{"`gunzip -c H2494_4.ps.gz"} } 
{\epsfbox[56 78 459 482]{"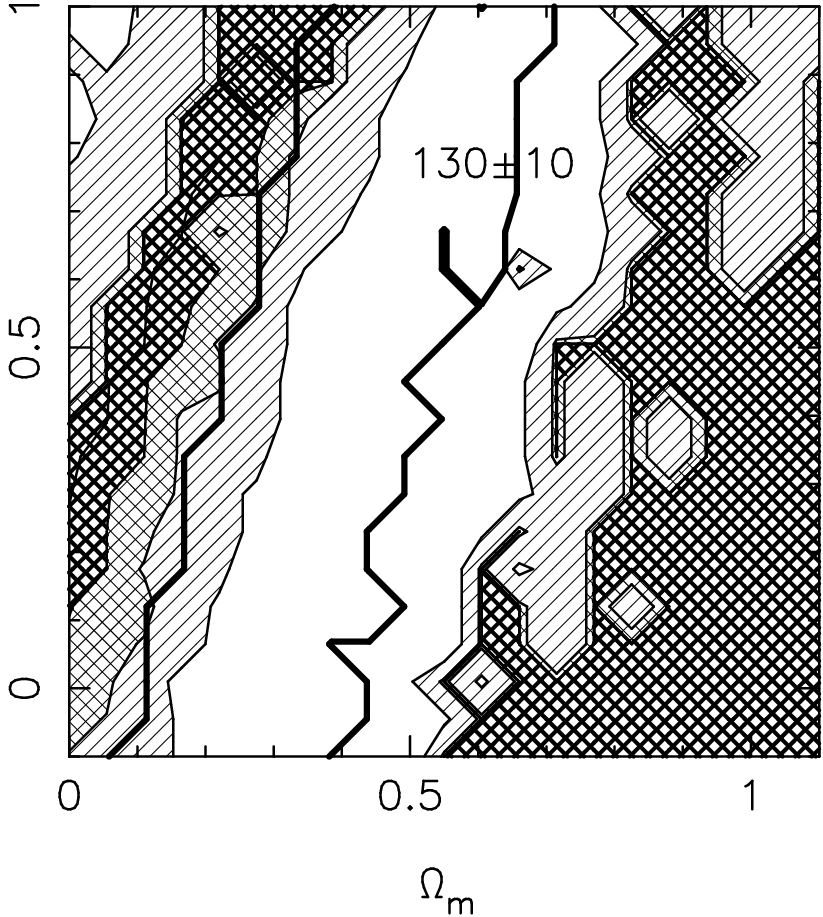"}}{}   
} 
}
\rule{0ex}{3mm}\par 
\caption[]{
Confidence intervals for rejecting the presence of a peak at 
$\llss\pm\Delta\llss$ for various hypotheses on $(\Omm,\Omega_\Lambda)$. 
Rejection levels are 
$1-P >$0\% (white), 
$1-P >$68\% (hatched), 
$1-P >$95\% (light cross-hatched) and 
$1-P >$99.7\% (heavy cross-hatched). 
The $1-P=68\%$ contour for testing a peak at 
$\llss/2\pm\Delta\llss/2$ test is shown in bold. 
For each pair $(\Omm,\Omega_\Lambda)$, 
the peak position is estimated as the value $\rmax$ 
for which $\xi(r)$ is maximum in 
$100 \le r \le 300${\hMpc}. 
(For the $\llss/2$ peak, $40 \le r \le 100${\hMpc} is searched). 
The measurement uncertainty in the $f\llss$ peak is estimated as 
$\Delta\rmax= f\Delta\llss$ (for $f=0.5,1$). 
The probability of finding $\rmax$ close to $f\llss$ 
assumes Gaussian errors, i.e. 
$P_f\equiv \mbox{\rm erfc}[ | \rmax-f\llss | /(\sqrt{2}\sigma) ]$, 
where $\sigma^2= (\Delta\rmax)^2+(f\Delta\llss)^2$. 
} 
\label{f-corrnb} 
\end{figure} 
} 

\def\fdegen{ 
\begin{figure} 
\centering 
\rule{0ex}{4ex} \par 
\dosingle 
{\epsfxsize=40mm %
\zzz{\epsfbox[56 68 459 452]{"`gunzip -c H2494_5a.ps.gz"} }{\epsfbox[56 68 459 452]{"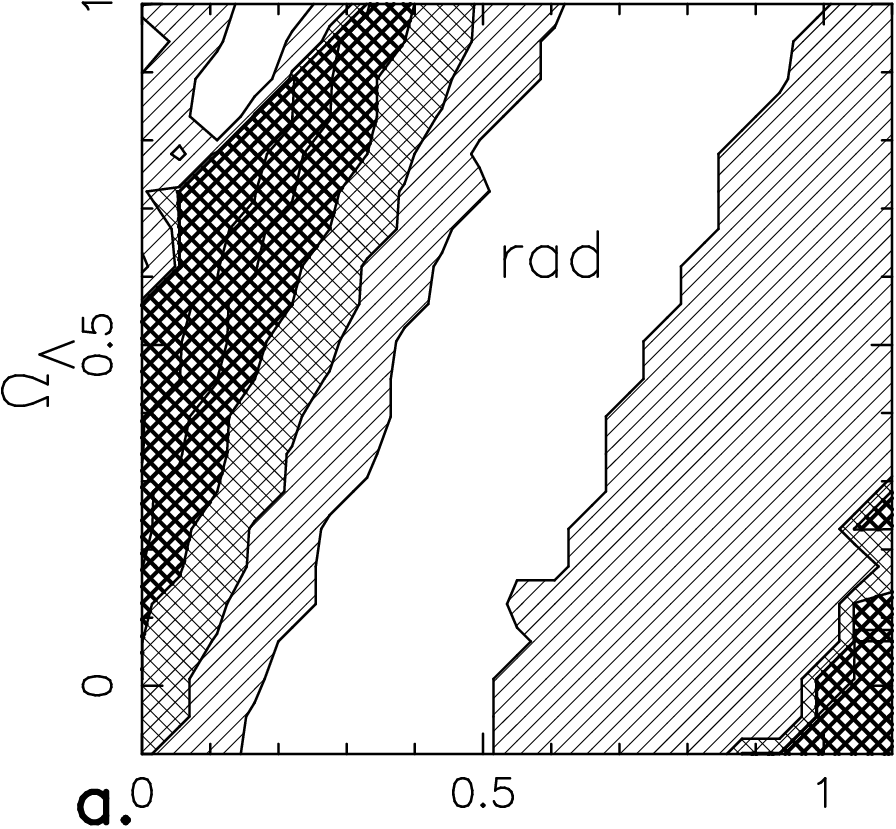"}}{}} 
{\epsfxsize=40mm %
\zzz{\epsfbox[-360 28 43 29]{"`gunzip -c H2494_5b.ps.gz"} }{\epsfbox[-360 28 43 29]{"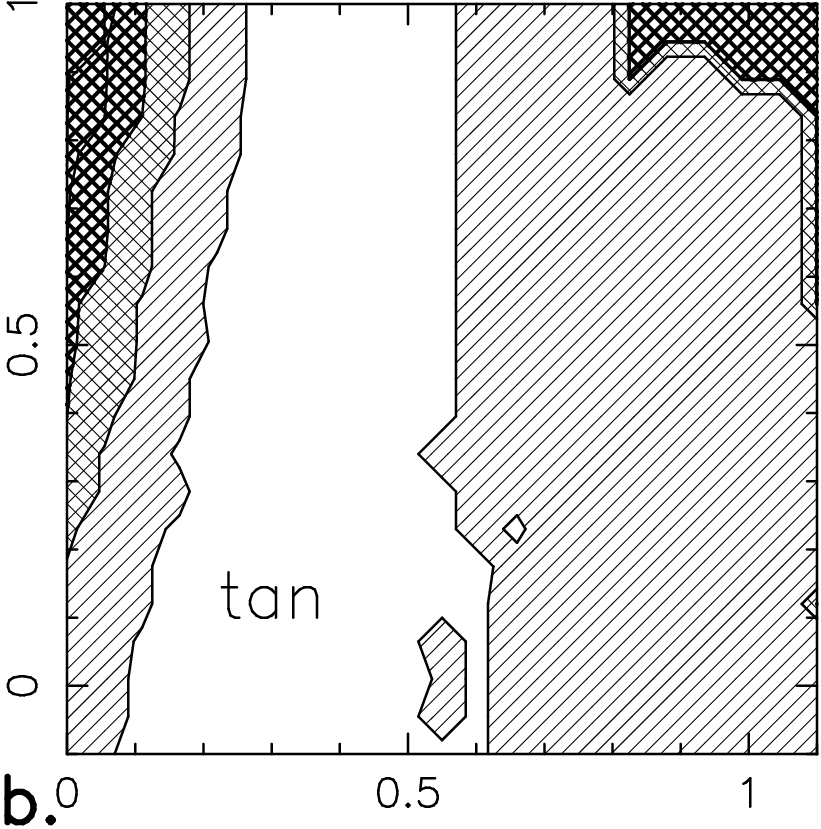"} }{}} 
{\epsfxsize=40mm %
\zzz{\epsfbox[56 78 459 472]{"`gunzip -c H2494_5c.ps.gz"} }{\epsfbox[56 78 459 472]{"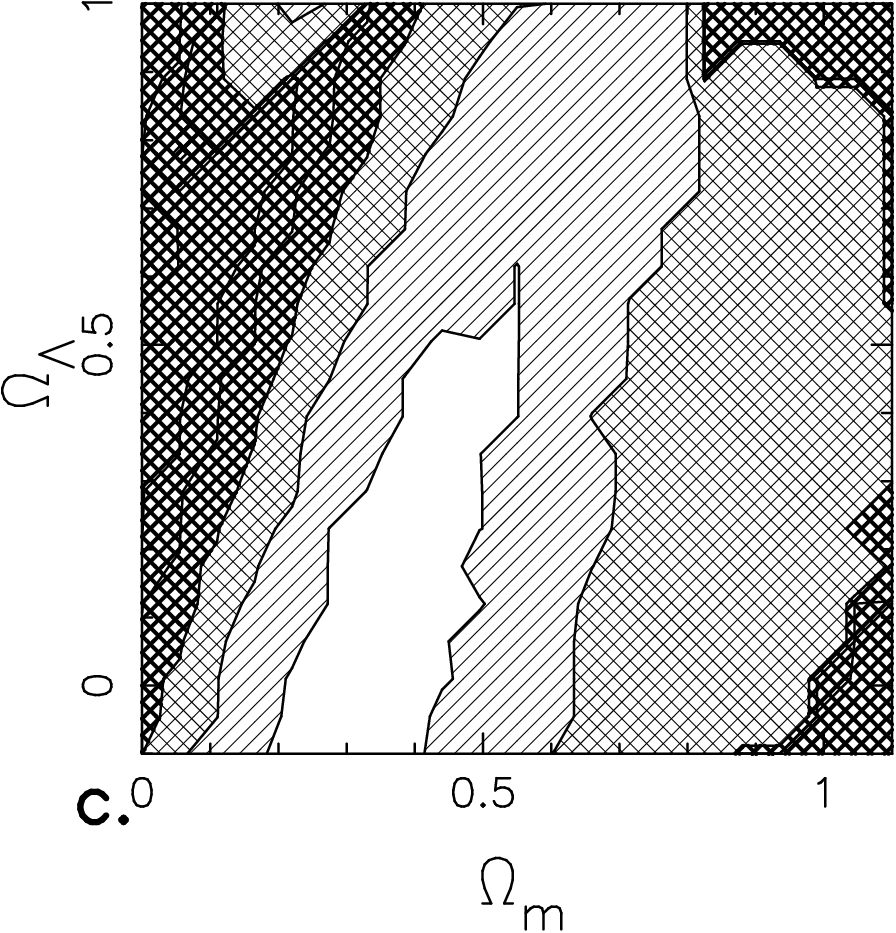"}}{}} 
{\epsfxsize=40mm %
\zzz{\epsfbox[-360 38 43 74]{"`gunzip -c H2494_5d.ps.gz"} }{\epsfbox[-360 38 43 74]{"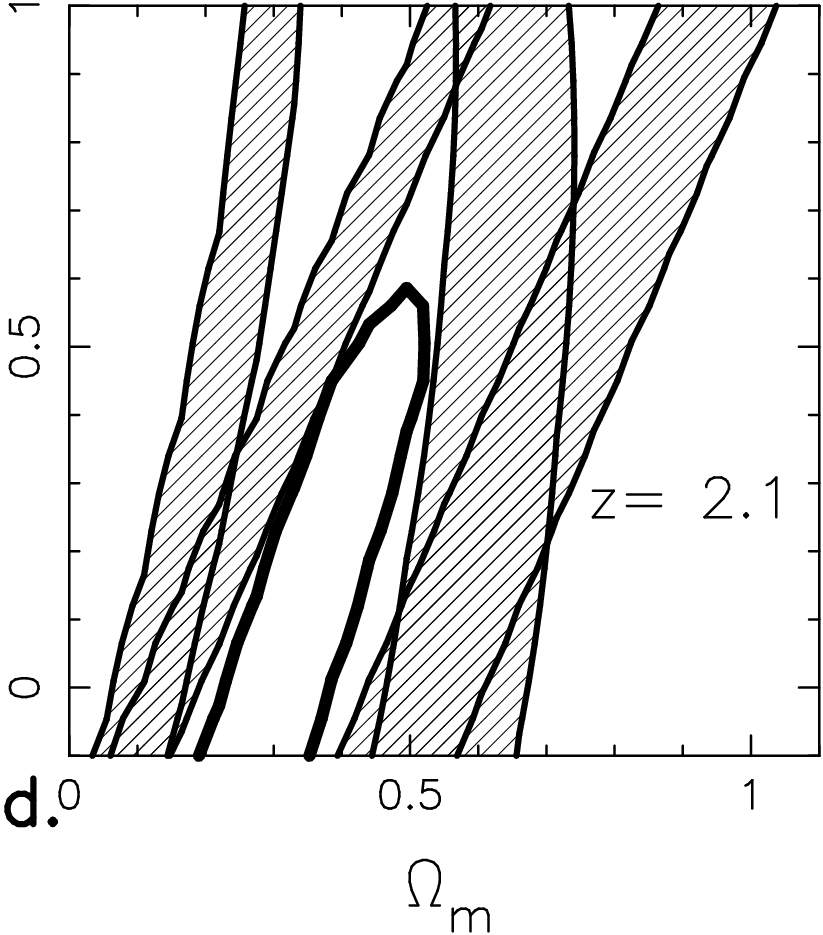"}}{}} 
::::
\dodouble
{\epsfxsize=40mm %
\zzz{\epsfbox[56 68 459 452]{"`gunzip -c H2494_5a.ps.gz"} }{\epsfbox[56 68 459 452]{"H2494_5a.ps"} }{}} 
{\epsfxsize=40mm %
\zzz{\epsfbox[-360 8 43 9]{"`gunzip -c H2494_5b.ps.gz"} }{\epsfbox[-360 8 43 9]{"H2494_5b.ps"}}{}} 
{\epsfxsize=40mm %
\zzz{\epsfbox[56 78 459 472]{"`gunzip -c H2494_5c.ps.gz"} }{\epsfbox[56 78 459 472]{"H2494_5c.ps"}}{}} 
{\epsfxsize=40mm %
\zzz{\epsfbox[-360 18 43 54]{"`gunzip -c H2494_5d.ps.gz"} }{\epsfbox[-360 18 43 54]{"H2494_5d.ps"}}{}} 
::::
\caption[]{ 
Partial lifting of the $\Omm-\Omega_\Lambda$ degeneracy. 
Confidence intervals are as for Fig.~\protect\ref{f-corrnb} for  
{\bf a.} the radial ($1-\Prad_1$), 
{\bf b.} tangential ($1-\Ptan_1$) and 
{\bf c.} combined ($1-\Prad_1 \Ptan_1$) constraints. 
{\bf d.} A simple model, for which the $95\% > 1-P > 68\%$ contours for 
radial (straight contours) and tangential (curved, nearly vertical contours) 
constraints showing near linear degeneracy are shaded as 
before (hatched regions); 
and the combined $68\%> 1-P> 0\%$ contour is the half-ellipse-like 
shape nested between these (heavy contour). 
{\bf a.-c.} The radial and tangential tests are performed for 
the $\llss\pm\Delta\llss$ peak as in Fig.~\protect\ref{f-corrnb}, 
except that only pairs oriented within $30\protect${\mydeg} of the radial 
and tangential directions (respectively) are included in 
calculation of $\xi(r)$. 
Both angular subsamples over $1.8 \le z < 2.4$ are used. 
The radial and tangential tests are assumed to be statistically
independent.
{\bf d.} For the model, the redshift interval 
$\Delta z$ and the angle $\theta$ corresponding to $\llss$,  
assuming that $(\Omm=0.3,\Omega_\Lambda=0.0)$ and that $z=2.1$, 
are calculated. 
For each pair $(\Omm,\Omega_\Lambda)$, the radial and tangential 
distance intervals implied by $\Delta z$ and 
$\theta$ are calculated, 
ignoring the initial assumption about $\Omm$ and $\Omega_\Lambda$. 
These are treated as two 
independent `experiments', and Gaussian probabilities 
$\Prad, \Ptan$ of observing these values, given 
$\llss,$ $\Delta\llss$, $\Omm$ and $\Omega_\Lambda$ 
are calculated as before. 
The combined rejection is $1- \Prad \Ptan$. 
} 
\label{f-degen} 
\end{figure} 
} 

\begin{document} 
\thesaurus{02 
(12.03.3; 
12.03.4; 
11.17.3; 
12.04.3; 
12.12.1; 
05.18.1) 
}

\title{
Lifting cosmic degeneracy within a single quasar survey} 

\author{B. F. Roukema\inst{1} \and 
G.~A.~Mamon\inst{2,3}}

\titlerunning{Lifting cosmic degeneracy} 
\authorrunning{Roukema \& Mamon} 
\date{\frtoday}

\institute{Inter-University Centre for 
Astronomy and Astrophysics, 
    Post Bag 4, Ganeshkhind, Pune, 411 007, 
India {\em (boud@iucaa.ernet.in)} \and
Institut d'Astrophysique de Paris 
(CNRS UPR 341), 
98bis Bd Arago, F-75014 Paris, France {\em (gam@iap.fr)} \and
DAEC (CNRS UMR 8631), 
Observatoire de Paris-Meudon, 5 place Jules Janssen, F-92195 Meudon Cedex, 
France}

\offprints{G.~A.~Mamon}

\maketitle

\begin{abstract}
In the almost Friedmann-Lema\^{\i}tre 
model of the Universe, the density parameter, $\Omm$, and the 
cosmological constant, $\Omega_\Lambda$, 
measure curvature. Several linearly degenerate relations between these 
two parameters have recently been measured. 
Here, large scale structure 
correlations at $\sim 100-150${\hMpc} are found in the comoving 
three-dimensional separations of redshift $z\approx 2$ 
quasars. These function as a comoving standard rod 
of length $\llss\approx 130\pm10${\hMpc}. 
A local maximum in the correlation function at 
$\approx \llss/2$ also appears to be significant. 
By combining separate 
radial and tangential standard ruler analyses, 
the lifting of the $\Omm-\Omega_\Lambda$ 
linear degeneracy within a single data set 
is demonstrated for the first time. 
\end{abstract}


\fwedge

\section{Introduction}

In standard cosmology \citep{Wein72}, space is 
a 3-manifold \citep{Schw00,LR99} of nearly constant curvature, i.e. space 
is approximately locally homogeneous. 
Geometric ways of measuring average curvature include the use of 
phenomena of intrinsically fixed brightness or length scale,
i.e. of standard candles \citep{SCP9812,HzS98}
and standard 
rulers [e.g. \citet{Mo92,BJ99,RM00a}; note also the microwave background
angular statistical estimates 
\citep{Boom00a,Maxima00a} 
which can loosely speaking be thought of
as `theoretical' standard rulers],
but have previously been found to lead to degeneracy in the 
$\Omm-\Omega_\Lambda$ plane (e.g. \citealt{Linew98}). 
However, inhomogeneities (perturbations) in density exist and 
can be statistically represented 
by a Fourier power spectrum, and are believed to gravitationally collapse 
and form objects such as galaxies and clusters of galaxies. 
Use of a characteristic feature of this spectrum at a scale 
$\gg 10${\hMpc}, the size of the largest bound structures, 
should provide a comoving standard ruler 
for constraining the local geometrical parameters $(\Omm,\Omega_\Lambda)$. 

Many observations of both galaxies and superclusters of galaxies 
indicate that the maximum in the power spectrum 
is peaked at a wavenumber $2\pi/\llss$, where 
$\llss\pm\Delta\llss 
\approx130\pm10${\hMpc} \citep{Bro90,GazB98,Einasto97nat,Deng96} 
(comoving length units).
Since this standard ruler should be valid 
independently of orientation, the different $\Omm-\Omega_\Lambda$ 
degeneracies implied in the radial and tangential applications of 
the ruler  
should enable lifting of the degeneracy within a single data set, 
providing a potentially more powerful ruler than previous standard 
rulers or standard candles. 

It should be noted that while the existence of a broad maximum
in the power spectrum is uncontroversial, 
not all observational analyses agree
on whether or not there is a sharp feature in the power spectrum
in this region in addition to the broad maximum, and there is not
yet any clear agreement on the characteristic scale of the broad maximum.
For example, on one hand, 
\citeauthor{Einasto97nat}'s (1997) analysis of superclusters
suggests a sharp peak at $ k=2\pi/L \sim 0.05${\hMpcinv}. 
But, on the other hand, while in the low redshift IRAS PSCz 
(point source catalogue redshift) survey \citep{Suth99}, 
there is, at least, a {\em broad} maximum at 
around $0.02${\hMpcinv}~$\la k=2\pi/L \la 0.04${\hMpcinv} 
(fig.~1 of \citealt{Suth99}), i.e. 
$320${\hMpc}~$\ga L \ga 160${\hMpc}, 
there is no obvious {\em sharp} feature in this region.  
Nevertheless, there is a $2\sigma$ significant sharp peak 
(see fig.~1 and comment in section 6 of \citealt{Suth99})
which lies at $0.07${\hMpcinv}~$\la k \la 0.10${\hMpcinv},
i.e. $90${\hMpc}~$\ga L \ga 60${\hMpc} in the PSCz. 

Possible reasons why the $\llss=130${\hMpc} feature found by 
other authors, if real,
might have been missed in the \citet{Suth99} analysis
include
\begin{list}{(\roman{enumi})}{\usecounter{enumi}}
\item use of a different population (lacking in early type galaxies),
\item use of too large a bin size, i.e. too much smoothing,
\item redshift distortion (velocity dispersion at small scales, 
smooth infall at larger scales), though this is discussed briefly
in the paragraph preceding Sect. 3.1 of \citet{Suth99},
\item assumption of zero cosmological constant [e.g. if 
($\Omm=0.3,\Omega_\Lambda=0.7$), then the length scale to 
$cz=45000\,$km/s is underestimated by 7\%],
\item assumption of zero curvature by use of a power spectrum analysis.
\end{list}

The method of \citet{RM00a} did {\em not} assume the
$\llss$ feature to be sharp, though a broad feature would obviously have given
a less significant (or maybe an insignificant) signal.
In the present analysis, a reasonably sharp (but low amplitude) feature 
consistent with $\llss\sim 130${\hMpc} is found.
A secondary feature
consistent with the $k \sim 0.08${\hMpc} feature of 
\citet{Suth99} is also found, but is not studied in detail.

Physics which could potentially be investigated in order to
explain the feature at $\llss$ includes 
acoustic oscillations in the baryon-photon fluid before last scattering, 
in high baryon density models \citep{Eisen98a,MeikWP98,Peeb99b}, 
and features from Planck epoch physics which transfer to oscillations 
in the post-inflation power spectrum, for weakly coupled 
scalar field driven inflationary models \citep{MB00b}. 
At high redshift, the $\llss=130\pm10${\hMpc} feature [``distance'' 
means comoving proper 
distance \citep{Wein72} throughout this paper] has been 
detected among quasars \citep{RM00a,Deng94} 
and Lyman-break galaxies \citep{BJ99}. 
 
Most applications of standard candles or standard rulers exploit either 
the radial redshift-distance relation \citep{BJ99} 
or the tangential redshift-distance-angle 
relation \citep{SCP9812,HzS98,RM00a,Boom00a,Maxima00a}, but not both 
simultaneously. 

\citet{AP79} suggested the idea of using both constraints simultaneously,
and suggested applying it at quasi-linear
or non-linear scales, i.e. $r \la 10${\hMpc}, but did not discuss
how to lift the degeneracy in the two curvature 
parameters $(\Omm,\Omega_\Lambda)$ which remains after using the local
isotropy constraint, though they did suggest a theoretical method
for separating out some of the 
peculiar velocity effects which are important at
these small scales.
\citet{Phil94}, \citet{MSuto96}, \citet*{Ball96} 
and \citet{Popowski98} followed up this idea, demonstrating 
specific formulae and calculations regarding quasar pairs and the two-point
auto-correlation functions of galaxies and quasars, including separation
of local isotropy (`sphericity') and some of the 
peculiar velocity effects. 

However, by using a standard ruler in the linear regime, i.e. by using 
a feature at $\llss\approx 130${\hMpc}, peculiar velocity effects become
negligible, and the inability of this scale to evolve in a Hubble 
time provides an additional constraint in the 
$(\Omm,\Omega_\Lambda)$ plane. 
For the $r \la 10${\hMpc} auto-correlation function, the peculiar
velocity effects are certainly important, 
and evolution in the length scale must be contended with, 
for example by model-dependent assumptions.

\tdefnsample

\fcorrna

\fcorrnab

\fcorrnb

\section{Observational Analysis and Discussion}

In a previous analysis \citep{RM00a} of a deep, dense, homogeneous quasar 
survey \citep{IovCS96}, 
only the tangential relation was used, 
to ensure that observational selection effects well known to cause 
non-cosmological periodicities in redshifts \citep{Scott91} 
could not bias the result. 
In the present analysis of the high grade quasar candidate catalogue
(Table~\ref{t-defnsample}, 
Fig.~\ref{f-wedge}), 
the technique of ``redshift scrambling'' 
(see Fig.~\ref{f-corrna} caption) 
is used to enable use of 
three-dimensional information in a way that avoids redshift 
selection effects. Since the redshifts used in the random 
and observational catalogues consist of exactly the same set of 
numbers, any redshift selection effects, which are independent of 
angle, should statistically cancel out 
\citep{Osmer81} in calculation 
of the correlation function $\xi(r)$ \citep{GroP77}. 
Since some real signal could also cancel out, in principle, 
this implies a conservative estimate of $\xi(r)$, i.e. a lower limit
to $|\xi(r)|$.

\subsection{Local maxima in the 3-dimensional correlation function}
\label{s-corr3d}

Figs~\ref{f-corrna} and 
\ref{f-corrnab}  show that, for reasonable values of 
$(\Omm,\Omega_\Lambda)$, a local maximum in the correlation 
function consistent with $\llss=130\pm10${\hMpc} 
is clearly present. By contrast, an 
$(\Omm=1,\Omega_\Lambda=0)$ universe would require this local
maximum to occur at $L\approx 100${\hMpc}, in contradiction with
the low redshift estimates of $\llss$.
A correlation function 
consistent with the standard \citep{GroP77} galaxy-galaxy correlation function 
$\xi(r) \approx (r/5${\hMpc}$)^{-1.8}$ is also present 
for $ r \la 40${\hMpc}.

What is the significance 
of the $\llss$ peak? This depends on where the zero level of 
correlation lies. In correlation function estimates where both 
sample and correlation are small, the problem of only 
having a finite volume often requires a correction known as 
the integral constraint \citep{GroP77}, which most often increases 
the precorrected values of $\xi$. Making an 
integral constraint correction usually requires assumptions on the 
intrinsic shape of $\xi$. To avoid these assumptions, 
it is more prudent just to quantify 
the peak as a local maximum \citep{Deng94,RM00a}. 
For a maximum at $\rmax$ consistent with a peak at 
$f\llss$, where $f=1$, define $\xi^r,$ $\xi_-$ and $\xi_+$ as 
the maximum value and the first minima below and above 
$\rmax$ respectively, and 
take the maximum value of $\sigma_{\left<\xi\right>}(r)$ for 
$r\in [\rmax-f\Delta\llss,\rmax+f\Delta\llss]$, 
where $\Delta\llss=10${\hMpc}. Then, 
\begin{equation} 
(S/N)_f \equiv { \xi^r - (\xi_- + \xi_+)/2 \over 
\max\{\sigma_{\left<\xi\right>}(r)\} } 
\label{e-defnsn} 
\end{equation} 
yields 
$(S/N)_1=3.2$ for $(\Omm, \Omega_\Lambda)=$ $(0.4, 0.6)$. 
 
\tprobsim

Fig.~\ref{f-corrnb} shows that an automatic search for this 
peak, using a simple and robust method, i.e. using the value of 
$r$ for which $\xi(r)$ is maximum over a very large interval in $r$, 
yields an approximately linear confidence band in the 
$(\Omm,\Omega_\Lambda)$ plane. Since this band is consistent with 
kinematical \citep{CYE97,Mam93} 
and baryonic fraction \citep{WNEF93,HMam94} 
constraints for clusters and groups of galaxies, 
though to slightly higher $\Omm$ values than were found 
in the purely tangential analysis of the present survey \citep{RM00a}, the 
coincidence would be surprising if it were due to noise or 
systematic effects. 
 
Moreover, what appear to be peaks at 
$\llss/2\pm\Delta\llss/2$ and near $2\llss\pm2\Delta\llss$ are 
present, though to lower significance, with 
$(S/N)_{0.5}=2.3$ and $(S/N)_{2}=1.2$ respectively, 
for $(\Omm, \Omega_\Lambda)=$ $(0.4, 0.6)$.
Could any of the three peaks be induced by noise which has 
common statistical properties among all the four subsets, 
either due to shot noise or 
selection effects? Redshift selection effects 
have been removed by the use of $z$-scrambling. Angular selection 
effects may be present at a small level 
(see section 3.4 of  \citealt{RM00a}), 
but are more likely to decrease the amplitude of any signal rather 
than introduce false correlations which mimic the signal found. 
Moreover, the convergence of the 
separate tangential and radial analyses below 
(\SS\ref{s-corrtanrad}) suggest that the effects of angular selection
are weak.
 
To test the properties of shot noise, 
random simulations were performed as before but substituted 
for the data. The probabilities that maxima can occur as close to 
and of at least the same signal-to-noise 
ratio as the observed values can be defined 
\begin{eqnarray} 
\Psim_f &\equiv &
P{\left[ 
{\rule[-0.3ex]{0ex}{3ex}\;} 
|\rsim
-f\llss| \le|\robs-f\llss|^{\mbox{\rm \ }}  
 \right.}  \anddd \nonumber \\
&& \left. (S/N)_f^{\mbox{\rm sim}} \ge (S/N)_f^{\mbox{\rm obs}} 
{\;\rule[-0.3ex]{0ex}{3ex}} 
\right], 
\label{e-defnpsim} 
\end{eqnarray} 
where `sim' and `obs' indicate simulations and observations 
respectively, and the intervals in $r$ are as above. 
Since one might suspect a single noise feature to cause, say, simultaneous 
features at two or three of the positions, mimicking the 
observed signal, the probabilities of finding 
the peaks might not be independent, e.g. 
$\Psim_{f_1,f_2} \approx \Psim_{f_1} \approx \Psim_{f_2} 
\gg \Psim_{f_1} \Psim_{f_2}$ for $f_1, f_2 \in \{0.5,1,2\}$ 
is in principle possible. 
The results (Table~\ref{t-probsim}) show that the hypotheses of 
{\em either} of 
the $\llss/2$ or the $\llss$ peaks occurring by chance are each rejected 
to $1-\Psim > 97\%$, and that of {\em both} occurring simultaneously is 
rejected to $1-\Psim > 99.5\%$. The hypothesis of the  
the $2\llss$ peak occurring by shot noise cannot be significantly 
rejected. 
These values vary throughout the $(\Omm,\Omega_\Lambda)$ plane. 
 
\fdegen

Fig.~\ref{f-corrnb} shows that the $68\% > 1-P > 0\%$ confidence 
intervals for the $\llss/2$ and $\llss$ peaks are consistent. 
 
\subsection{Tangential versus radial correlations}
\label{s-corrtanrad}

A standard ruler should not depend on orientation. Can use 
of both radial and tangential information lift the degeneracy of 
$(\Omm,\Omega_\Lambda)$ constraints? 
To illustrate 
this, the full redshift interval $1.8 \le z< 2.4$ 
is used, but only pairs of objects within $30${\mydeg} of either the 
radial or tangential directions respectively are used. 
For the geometry of this survey, 
about 10\% of pairs are radial and 60\% of pairs are 
tangential according to this criterion. 
 
Fig.~\ref{f-degen} 
clearly shows, both observationally and theoretically, 
the difference in the slopes of the radial and tangential constraints 
at $z\approx 2$. 
A hyperbolic ($\Omm+\Omega_\Lambda-1<0$) universe is suggested 
by the 68\% confidence limit, though 
a flat universe with $\Omega_\Lambda = 1-\Omm =0.5$ 
is only rejected to $1-P\approx 80\%$ confidence, i.e. not significantly. 
The partially lifted degeneracy can be represented (at 68\% confidence) as 
\begin{equation} 
\Omm= (0.30\pm0.04)\Omega_\Lambda +(0.30\pm0.11), 
\;\;\Omega_\Lambda < 0.5. 
\label{e-newresult} 
\end{equation} 
This suggests a somewhat higher matter density and lower cosmological
constant than other recent results.

\section{Conclusion}

The confirmation of the $\llss$ peak and the partial lifting 
of the $\Omm-\Omega_\Lambda$ degeneracy show that ongoing 
and future large quasar surveys [in particular the 2 Degree Field
Quasar Survey \citep{Boy00a} 
and the Sloan Digital Sky Survey quasar sample (e.g. \citealt{Fan00})]
will have a much more powerful tool for 
local geometrical constraints than was previously thought. 
While local isotropy of the $r \la 10${\hMpc} scale
correlation function can in principle be used as a local geometrical
constraint, a standard ruler at $\llss\approx130${\hMpc} has 
the advantages (i) of being little affected by peculiar velocities,  
and (ii) of occurring well into the linear regime where 
evolution within a Hubble time is unlikely.

Moreover, the detection of the $\llss/2$ 
peak (cf fig.~6 of  \citealt{TadE96}; fig.~1 of 
\citealt{Suth99}; fig.~3 of \citealt{Mo92}) implies that 
both peaks might either be 
signs of high baryon density \citep{Eisen98a,MeikWP98,Peeb99b} or of 
pre-inflationary physics \citep{MB00b}, enabling constraints to 
be put on these. 
For increased confidence in this method, more precise low redshift 
constraints on large scale structure features near $\llss\approx 130${\hMpc} 
will be highly desirable. Results from the 
2 Degree Field Galaxy Redshift Survey 
(e.g. \citealt{2dFGRS99}),
the Sloan Digital Sky Survey galaxy sample 
\citep{York00}, and the 6 Degree Field
galaxy survey (e.g. \citealt{Mam98}) 
may help for these low redshift calibrations.


 \section*{Acknowledgments} 
We thank Emmanuel Bertin, St\'ephane Colombi, Georges Maignan, S. Sridhar 
and the referee, Pat Osmer, 
for useful comments. B.F.R. thanks 
the Institut d'Astrophysique de Paris, CNRS and 
DARC, Observatoire de Paris, for their hospitality, 
and acknowledges 
the support of la Soci\'et\'e de Secours des Amis des Sciences. 
Data are available 
at {\em http://cdsweb.u-strasbg.fr/cgi-bin/Cat?J/A+AS/119/265}.


\end{document}